\documentclass[12pt]{article}

\usepackage[a4paper, top=2.5cm, bottom=2.5cm, left=2.2cm, right=2.2cm]{geometry}
\usepackage{amsmath, amssymb}
\usepackage[utf8]{inputenc}
\usepackage{bm}
\usepackage{graphicx}
\usepackage{tabularx}
\usepackage{enumitem}
\usepackage[numbers,sort&compress]{natbib}
\usepackage{color}
\usepackage{subcaption}
\usepackage{hyperref}
\usepackage{booktabs}
\usepackage{array}
\usepackage{upgreek}
\usepackage{appendix}
\usepackage[justification=raggedright,singlelinecheck=false,format=plain,labelsep=colon]{caption}

\begin{document}
	
	\title{Massive Black Hole formation in proto-stellar clusters via early gas accretion}
	
	\author{Zacharias Roupas$^{1,2}$\\[0.5em]
		\small $^{1}$ Dipartimento di Fisica ``G. Occhialini'',
		Universit\'a degli Studi di Milano-Bicocca,\\
		\small Piazza della Scienza 3, 20126 Milano, Italy\\[0.3em]
		\small $^{2}$ Istituto Nazionale di Fisica Nucleare (INFN), Sezione di Milano-Bicocca,\\
		\small Piazza della Scienza 3, 20126 Milano, Italy\\[0.5em]
		\small Correspondence: \href{mailto:zacharias.roupas@unimib.it}{zacharias.roupas@unimib.it}}
	
	\date{}
	
	\maketitle
	
	\begin{abstract}
		We review our semi-analytic model of stellar black hole (BH) mass growth by gas accretion in gas-rich stellar clusters during their birthstage within the first $\sim 10\,{\rm Myr}$ after the first stellar formation event. Such proto-stellar clusters are massive and compact, with typical masses $\sim 10^6\,{\rm M}_\odot$ and sizes $\sim 1\,{\rm pc}$, suggested by recent James Webb Space Telescope (JWST) observations. We find that the BH masses are shifted by the end of gas depletion to values within and above the BH mass gap, well within the range of components of the recent gravitational-wave (GW) signal GW231123, and up to masses $\sim 10^3\,{\rm M}_\odot$.
	\end{abstract}

	\section{Introduction}
	
	In Ref. \cite{2025A&A...702A.208R}, we proposed a channel for the BH mass growth via gas accretion in proto-stellar clusters to values within the upper BH mass gap, induced by the physics of a pair-instability supernova (PISN) \cite{2016A&A...594A..97B,2017MNRAS.470.4739S,2021ApJ...912L..31W}, $60\,{\rm M}_\odot \lesssim m_{\rm BH} \lesssim 130\,{\rm M}_\odot$, and up to the intermediate-massive-black-hole (IMBH) regime of $m_{\rm BH} \sim 10^3\,{\rm M}_\odot$. We review briefly the model and main results here.
	
	Gas-rich, compact, low-metallicity, massive stellar clusters provide ideal conditions for the mass growth of stellar BHs via accretion. Mass segregation, which drives the massive stellar progenitors of BHs toward the dense center of the cluster \cite{Spitzer_1987degc}, is accelerated in the presence of ambient gas at timescales $\lesssim 1\,{\rm Myr}$ \cite{2014MNRAS.441..919L}. In addition, a lower-metallicity environment weakens stellar winds, allowing for longer gas-depletion timescales in a sufficiently compact cluster, even if they allow for energetic PISN to occur (which inject significant energy to the ambient gas) \cite{2025A&A...702A.208R}. 
	
	Recent JWST observations revealed five proto-stellar cluster candidates in the Cosmic Gems arc galaxy at high redshift, $z = 10.2$, \cite{2024Natur.632..513A}, likely progenitors of globular clusters \cite{2015MNRAS.454.1658K,2017MNRAS.469L..63R,2018ApJ...869..119E}. Their typical masses are $\sim 10^6\,{\rm M}_\odot$ and half-light radii $\sim 1\, {\rm pc}$, providing observationally motivated cluster parameters for model-building.
	Our semi-analytic framework of stellar-BH growth in such environments \cite{2025A&A...702A.208R} accounts for gas depletion by stellar feedback, gas accretion, distinct BH formation times, dynamical friction, and stochastic kicks induced by stellar encounters and gas turbulence.
	
	Identifying the origin of heavy BHs $\sim 10^2-10^3\,{\rm M}_\odot$ is a timely, active area of research  
	\cite{2017PhRvD..95l4046G,2019A&A...632L...8R,2020PhRvD.102d3002B,2020ApJ...904L..13R,Kimball_2020,2020ApJ...903L..21S,2021MNRAS.505..339M,2022MNRAS.512..884R,2023MNRAS.526..429A,2024A&A...688A.148T,2024Sci...384.1488F,2024AJ....167..191P,2025PhRvD.111f3039B,2025ApJ...988...15L,2025arXiv250808558B,2025ApJ...994L..37K,2025ApJ...993L..54G,2025ApJ...993L..30L}.
	The GW experiments, LIGO-Virgo-KAGRA Collaborations, have revealed a significant population of BHs within the upper BH mass gap \cite{2023PhRvX..13d1039A}. The most recent such GW signal, GW231123 \cite{2025ApJ...993L..25A}, involves components which are both heavy, $103_{-52}^{+20}\, {\rm M}_\odot$, $137_{-17}^{+22}\, {\rm M}_\odot$. Such values are naturally produced in our model. Furthermore, our model offers a pathway toward the formation of IMBH in the early Universe, providing the potential to explain the observed, intriguing, abundance and masses of Supermassive BHs (SMBH) at high redshifts \cite{2023ApJ...953L..29L, 2024NatAs...8..126B, 2024OJAp....7E..72R}. This open issue is also important for the Laser Interferometer Space Antenna (LISA) mission \cite{2023LRR....26....2A}. 
	
	In the next section we review our cluster and BH motion model.
	Section \ref{sec:results} presents our results, and we conclude in Section \ref{sec:conclusion}. 
	
	\section{Physical model}\label{sec:model}
	
	\subsection{Gas depletion timescale}
	
	In a gas-rich cluster, stellar winds and SN explosions inject energy and momentum to the ambient gas, which drive its expulsion out of the cluster. Stellar lifetimes determine both the time when SN occur, and the birth time of BHs from their stellar progenitors. Stellar formation in dense gas is completed rapidly ($\lesssim 1 {\rm Myr}$) \cite{2012MNRAS.420.1457H,2025ApJ...982..138O}, and is therefore justified to assume it occurs in a constant metallicity environment. Note also that stellar winds of solar-type intermediate-mass stars of $m < 8 \,{\rm M}_\odot$ contribute negligible energies within the depletion
	timescales $\sim 10\,{\rm Myr}$ (e.g. \cite{1999ApJS..123....3L}), and are therefore neglected. 
	We use phenomenological relations \cite{2025A&A...702A.208R} to model stellar winds, SN energy injection, and stellar lifetimes according to observational and numerical data. We estimate the depletion timescale $\tau$ as the moment when the accumulated injected energy becomes equal to the initial binding energy of the gas.

\begin{figure}[tbp]
		\centering
		\includegraphics[width=0.6\textwidth]{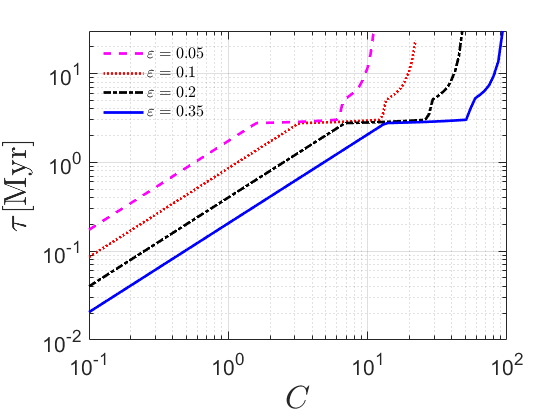}
		\caption{Depletion timescale, $\tau$, with respect to the compactness, $C$, of a gas-rich stellar cluster of any total mass at low metallicity for different possible star formation efficiencies, $\varepsilon$.}
		\label{fig:tau_dep}
	\end{figure}

	We find \cite{2025A&A...702A.208R} that the depletion timescale depends primarily on the compactness of the cluster (see also \cite{2016A&A...587A..53K})
	\begin{equation}
		C \equiv \frac{M/10^5{\rm M}_\odot}{r_{\rm h}} ,
	\end{equation}
	rather than on the cluster mass alone. This is because the binding energy scales as $U \propto C \cdot M$, while the total stellar feedback $E_{\rm feed} \propto M$ regardless of compactness. 
	In Fig.~\ref{fig:tau_dep}, we display $\tau$ as a function of compactness for several values of stellar formation $\varepsilon$, and for low metallicity $Z=0.01Z_{\odot}$. In \cite{2025A&A...702A.208R}, the reader can find additional figures for solar and sub-solar metallicities. 
	We observe a characteristic stall at 
	\begin{equation}\label{eq:tau_stall}
		\tau_{\rm stall} \approx 2.9{\rm Myr} 
	\end{equation} 
	which marks the onset of PISN. The range of compactness values used in our simulations, and which satisfy the ones of observed proto-stellar clusters in the Cosmic Gems arc galaxy, correspond to this stall timescale.
	
	\subsection{Stellar cluster and BH motion}

	We assume that the background gas and stellar densities follow Plummer profiles. We account for several important physical processes in the cluster: cluster expansion driven by gas mass loss, the greater spatial extent of the gas component relative to the stellar component, the gas temperature profile, and gas cooling over time.
	We adopt the stadard exponential law of gas depletion \cite{2007MNRAS.380.1589B}, 
	\begin{equation}
		\frac{d {M}_{\rm gas}(t)}{dt} = -\frac{M_{\rm gas}(t)}{\tau},
	\end{equation} 
	since the stellar feedback generates energy proportional to the cluster mass. Each run is terminated once $99\%$ of the initial gas mass has been depleted. 
	
	Our adopted initial BH mass function (BHMF) is the Salpeter profile \cite{2019ApJ...878L...1P}. The BH positions and velocities are sampled from the Brownian probability density of Ref. \cite{2021A&A...646A..20R}. Each BH is created at a distinct time based on the zero-age main sequence (ZAMS) progenitor mass. The correspondence of ZAMS progenitor to BH mass is applied according to Figure 2 of Ref. \cite{2017MNRAS.470.4739S}.  
	We classify the BHs on several types depending on the ZAMS progenitors mass: core-collapse-SN (CCSN), direct-collapse of low or high ZAMS mass, and pulsational-pair-instability-SN (PPISN) of low or high ZAMS mass. 
	We impose an upper cut-off on ZAMS mass at $230 {\rm M}_\odot$ and a cut-off on the upper initial BH mass at $55 {\rm M}_\odot$ as suggested by stellar evolution models \cite{2016A&A...594A..97B,2017MNRAS.470.4739S,2021ApJ...912L..31W}. This also allows us to isolate the BH gas accretion channel for populating the upper BH mass gap and quantify its sole contribution.
	Note, however, it has been argued that supermassive stars ($m_{\rm ZAMS} > 230{\rm M}_\odot$) could be created either by runaway collisions \cite{2006MNRAS.368..121F,2006MNRAS.368..141F,2023MNRAS.526..429A,2025A&A...704A.321V} or rapid gas accretion \cite{2012ApJ...756...93H,2015MNRAS.452..755S,2019A&A...632L...2H} in stellar clusters. Those stars could potentially generate BHs within the upper BH mass gap via direct collapse. The extent to which such supermassive stars contribute to our proposed BH growth channel warrants further investigation. Nevertheless, since they would only enhance the population of the BH mass gap, our analysis represents a conservative calculation of the BH mass function shift in proto-stellar clusters.
	
	We integrate the BHs equations of motion with a standard drift-kick scheme \cite{2025A&A...702A.208R}. The deterministic forces -- gravity and dynamical friction from the background-- were implemented using a standard Runge-Kutta method. 
	For the dynamical friction induced by the stellar component we use the standard Chandrasekhar formula \cite{1943ApJ....97..255C}. 
	The dynamical friction induced by gas requires modification to account for the hydrodynamic nature of the interaction as in Ref \cite{1999ApJ...513..252O}, while we also include the deceleration due to the mass increase \cite{2009ApJ...696.1798T,2011MNRAS.416.3177L,2014A&A...561A..84L,2019A&A...621L...1R}.
	
		\begin{figure}[tbp]
		\centering
		\includegraphics[width=0.6\textwidth]{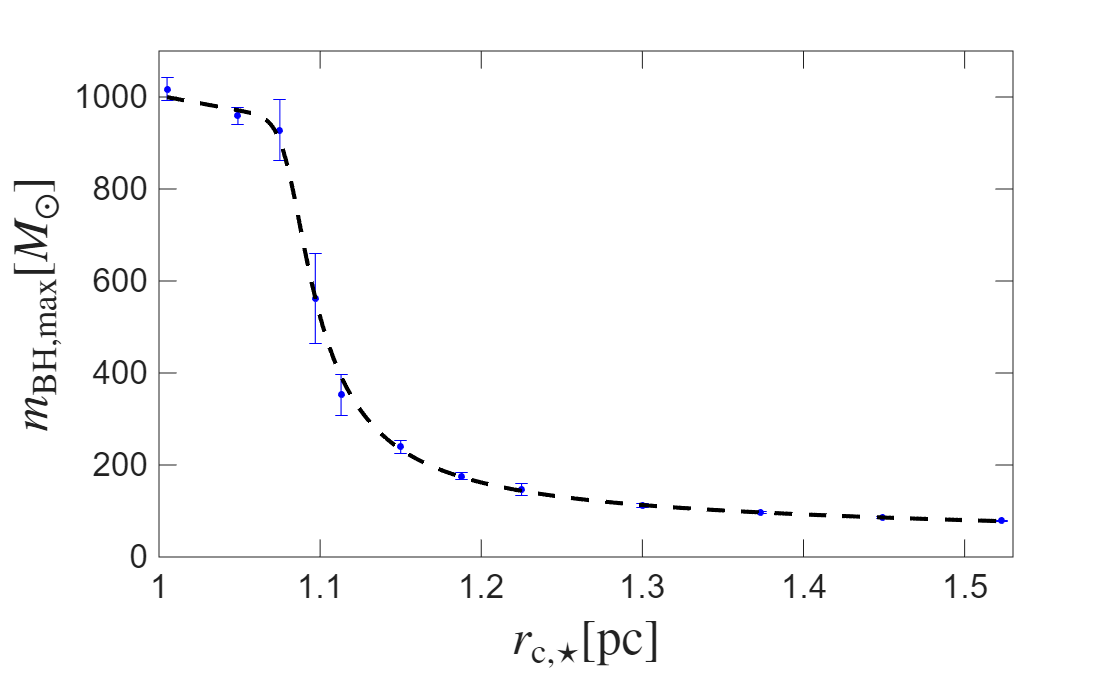}
		\caption{The sample maximum shifted individual BH mass in a stellar cluster with total stellar mass $M_\star = 10^6 {\rm M}_\odot$ and size $r_{{\rm c},\star} = 1{\rm pc}$, values representative of the observed Cosmic Gems proto-stellar clusters. The $x$-axis is the final, after gas depletion, stellar Plummer radius, to be compared to the half-light radius of the observed stellar clusters.}
		\label{fig:m_max_rh_Cosmic_Gems_M3e6}
	\end{figure}

	We applied at adaptive timesteps, stochastic velocity increments, due to both stellar perturbations and gas turbulence. 
	Those timestep values $\Delta t_{\rm stoch}(r_{\rm BH},t)$ are determined dynamically at the end of each previous step by  the minimum of the relevant physical timescales, locally for each BH. Those are the dynamical timescale, the crossing timescale, the dynamical friction timescale and the turbulence correlation timescale. We apply bounds $10^{-5}\,{\rm Myr} \leq \Delta t_{\rm stoch} \leq 10^{-1}\,{\rm Myr}$. Those are in effect never violated, since we get in practice that $\Delta t_{\rm stoch} \approx 10^{-4}-10^{-2}\, {\rm Myr}$ for any $r_{\rm BH}$ and $t$.
	The stochastic kicks on BHs due to stellar encounters are modelled as a white noise process (e.g. \cite{2021A&A...646A..20R}). Gas turbulence on the other hand is modeled as an Ornstein-Uhlenbeck (OU) process acting on the BH through a time-correlated turbulent acceleration field. This OU processs is a well-defined stochastic process with a finite autocorrelation timescale and can be used to excite turbulent motions in simulations \cite{1988CF.....16..257E,SCHMIDT2006353,2010A&A...512A..81F}. We integrate the OU stochastic differential equation directly to get the turbulent acceleration and the corresponding velocity kick. 
	
	We adopt an isotropic hot-type accretion, whose typical representative is Bondi-Hoyle accretion \cite{Merritt_book}.
	The spherically symmetric accretion rate is \cite{2002apa..book.....F}
	\begin{equation}\label{eq:dm}
		\dot{m}_{\rm feed} = \pi \rho_\text{gas} v_\text{rel} R_\text{acc}^2
		,\quad
		v_\text{rel} = \sqrt{v_\bullet^2 + c_s^2}.
	\end{equation} 
	The appropriate value of accretion radius, $R_{\rm acc}$ depends on the relative amount of gas and the radial position of the accretor \cite{2001MNRAS.324..573B}.
	If the gas dominates the cluster potential at a certain BH position, $r_{\bullet}$, the accretion rate is determined by a tidal-lobe accretion radius
	\cite{1971ARA&A...9..183P}.
	Otherwise, when gas is less dominant, the appropriate accretion radius is the Bondi-Hoyle radius \cite{2002apa..book.....F}.
	We assume the accretion radius to be the smaller of the two at any instant in time \cite{2001MNRAS.324..573B}
	adopting the most conservative assumption. 
	We impose a maximum threshold for accretion rate
	\begin{equation}\label{eq:dm_acc}
		\dot{m}_{\rm acc}(t) = \min\left\lbrace \dot{m}_{\rm feed}(t), 10 \dot{m}_{{\rm Edd},0} \right\rbrace
	\end{equation}
	where $
	\dot{m}_{{\rm Edd},0} = 4\pi G m_{\rm BH} m_p/\eta_0 \sigma_T c 
	$
	is the reference Eddington rate value for the radiative efficiency $\eta_0 = 0.057$. 
	This is a conservative cap based on the results of 
	\cite{2016MNRAS.459.3738I}. In practice, we find $0 < \dot{m}_{\rm feed}(t)/\dot{m}_{\rm Edd,0} < 10$, which satisfies the cap automatically.
	
	\section{Results}\label{sec:results}

\begin{figure}[tbp]
	\centering
	\begin{subfigure}[t]{0.48\textwidth}
	\centering
		 \raisebox{0.04cm}{\includegraphics[width=\textwidth]{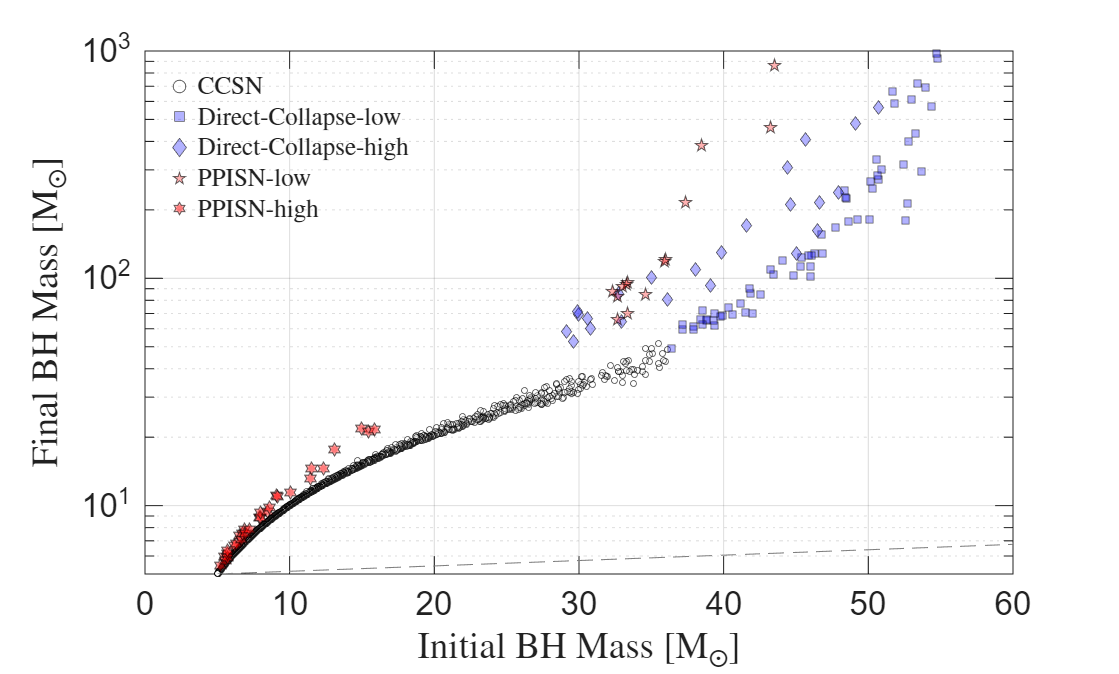}}
		\subcaption{}
		\label{fig:BHmass_scatter_M3.0e6_C41.7_eps0.35_Zlow}
	\end{subfigure}
	\begin{subfigure}[t]{0.48\textwidth}
	\centering
		\includegraphics[width=0.92\textwidth]{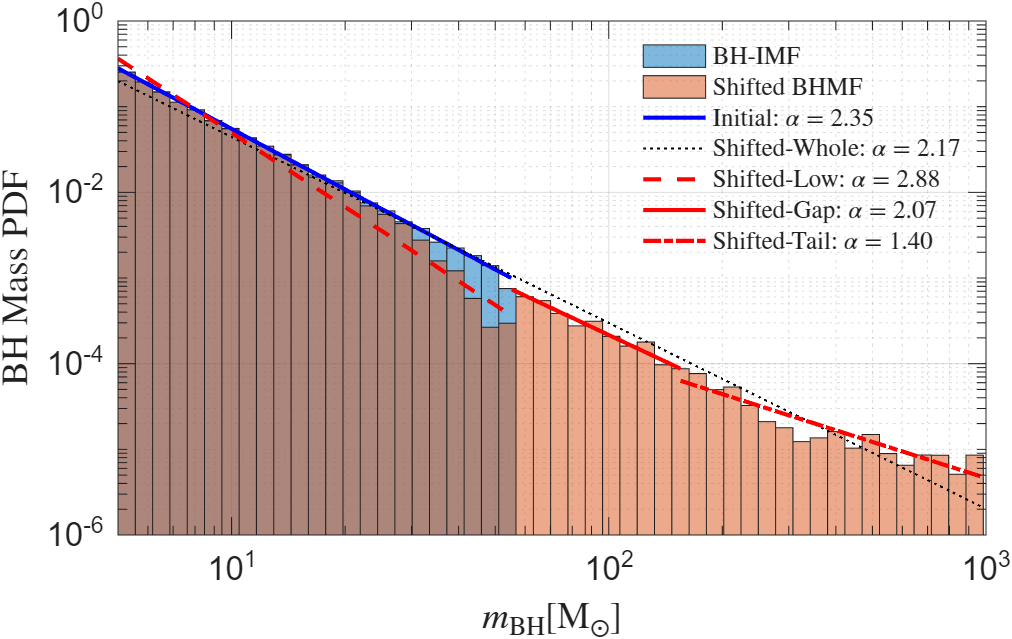}
		\subcaption{}
		\label{fig:BHMF_M3.0e6_C41.7_eps0.35_Zlow}
	\end{subfigure}
	\caption{Mass scatter diagram~(a), and BH mass function~(b), for a cluster
		with total stellar mass $M_\star = 10^6\,{\rm M}_\odot$ and final size
		$r_{{\rm c},\star} = 1\,{\rm pc}$, representing the observed Cosmic Gems
		proto-stellar clusters. In~(a) the data refers to a single typical
		simulation. In~(b) we use ten samples. The parameter $\alpha$ denotes
		the exponent of each histogram fit.}
	\label{fig:BHMF_cosmic_gems}
\end{figure}

	We model proto-stellar clusters, resembling the observed ones in the Cosmic Gems arc galaxy \cite{2024Natur.632..513A}. Their observed masses range $(1-4)\cdot 10^6{\rm M}_\odot$ and the half-light radii $(0.7 - 1.7) {\rm pc}$.
	We adopt $\varepsilon = 0.35$ representing the upper end $\sim 0.3$ of observed star formation efficiencies in embedded clusters \cite{2003ARA&A..41...57L,2007MNRAS.380.1589B}, appropriate for the extreme densities and low metallicity in such massive, compact proto-stellar clusters.
	We assume a low metallicity $0.01 Z_\odot$, in accordance with the low metallicity expected for the Cosmic Gems clusters \cite{2024Natur.632..513A}.
	
	We choose initial total mass of the gas-rich cluster $M = 3\cdot 10^6{\rm M}_\odot$, which means $M_{\star} = 10^6{\rm M}_\odot$ for the stellar component, that is the low-mass end of the observations. 
	In our investigation, we cover a range of initial half-mass radii $r_{\rm h} = 0.6-1{\rm pc}$ which result to stellar Plummer radii $r_{{\rm c},\star} = 0.9-1.6{\rm pc}$ because of expansion due to mass loss. For all these cluster sizes, the clusters have identical gas depletion timescales $\tau = 2.9 \,{\rm Myr}$,
	which implies that by $13{\rm Myr}$ the $99\%$ of the gas has been expelled when we terminate the simulation.
	
	We find that the upper BH mass gap is populated even for the most conservative cluster sizes. In Figure \ref{fig:m_max_rh_Cosmic_Gems_M3e6} we depict the maximum individual BH mass with respect to $r_{{\rm c},\star}$, to be compared with the observed half-light radius of Cosmic Gems clusters. 
	Remarkably, resulting stellar clusters with $r_{{\rm c},\star} \lesssim 1 \,{\rm pc}$ can host IMBH with mass $\sim 10^3\,{\rm M}_\odot$.
	In Figure \ref{fig:BHMF_cosmic_gems} we present the results of ten runs for this typical size.
	Our empirical general fit of the individual BH mass increase for all BHs, describing the trend of Figure \ref{fig:BHmass_scatter_M3.0e6_C41.7_eps0.35_Zlow}, is
	\begin{equation}\label{eq:dm_fit}
		\Delta m_{\rm BH} \approx \left(m_{{\rm BH},{\rm birth}}/25{\rm M}_\odot \right)^{7},	
	\end{equation}
	where the characteristic mass ranges $\sim (20-30)\,{\rm M}_\odot$ and the exponent $\sim 6-8$ for different cluster sizes.
	The BHMF shifts to higher masses (Figure \ref{fig:BHMF_M3.0e6_C41.7_eps0.35_Zlow}) populating the mass-gap with a negative power exponent $\simeq 2.1$ fitted within the mass-gap range. The power exponent of the tail $m_{\rm BH}>130{\rm M}_\odot$ is $1.4$.

	\section{Discussion and conclusion}\label{sec:conclusion}
	
	We have presented a semi-analytic model of BH mass growth in gas-rich proto-stellar clusters. Our analysis accounts for the  gas depletion timescale incorporating stellar lifetimes, stellar winds, and SN explosions, the generation of realistic BH populations with assigned progenitor masses, the input of BHs into the cluster according to their formation times, and stochastic perturbations from both gas and stellar components. 
	The BH mass growth is found to satisfy a power law $\Delta m = (m_{\rm BH}(0)/m_c)^\beta$ with parameter values which depend on cluster properties, and take typical values $m_c \sim 25$ and $\beta \sim 7$ for low metallicity clusters with masses $\sim 10^6\,{\rm M}_\odot$ and sizes $\sim 1\,{\rm pc}$. 
	Our analysis has direct observational implications. The observed Cosmic Gems proto-stellar clusters at high redshift likely experienced early accretion-driven BH mass function shifts to higher masses, generating $\gtrsim 25$ BHs with masses $10^2 {\rm M}_\odot \lesssim m_{\rm BH} \lesssim 10^3 {\rm M}_\odot$ within their first $\lesssim 10 {\rm Myr}$. 
	
	Our mechanism can provide further an explanation for some of the heavy, mass-gap BHs ($>60{\rm M}_\odot$) detected by LIGO-Virgo-KAGRA. 
	The key observable that may allow determine the origin of the BH growth mechanism is BH spin. 
	Particularly, the recent signal GW231123 \cite{2025ApJ...993L..25A} had masses $m_1 = 137_{-17}^{+22} \,{\rm M}_\odot$, $m_2 = 103_{-52}^{+20} \,{\rm M}_\odot$, and spins $a_{*,1} = 0.9_{-0.19}^{+0.10}$, $a_{*,2} = 0.8_{-0.51}^{+0.20}$. Let us discuss if our proposed mechanism can provide such high spins. Gas accretion spins up a Schwartzschild BH ($a=0$) up to an extreme Kerr BH ($a=1$) after having accreted mass $\sqrt{6}\,{\rm M}_{\odot}$ \cite{1970Natur.226...64B}. A BH that grows within the BH mass gap in our system accretes typically at a rate $2-10\,{\rm M}_\odot / {\rm Myr}$. This implies that it shall reach the maximum spin in a timescale $\lesssim 2\,{\rm Myr}$ that is less than the total depletion time-window of $\sim 10\,{\rm Myr}$. Thus, there is sufficient time for BHs to spin-up at high spin values comparable to GW231123. We are currently working on the modelling of BH spin-up and spin distribution, which will be reported elsewhere \cite{2026_ROUPAS}.
	
	This work opens several avenues for future investigation: 
	(i) the role of our proposed BHMF-shift mechanism in seeding high-redshift SMBH formation,
	(ii) IMBHs produced through this channel represent promising targets for GW detection by the future LISA mission;
	(iii) binary BHs which acrrete gas generate a GW background, which is important for GW missions, such as LISA;
	(iv) calculate the, possibly, distinctive spin signature of our mechanism and perform data analysis to distinguish from merger-formed BHs in current and future LIGO-Virgo-KAGRA GW data.
	
	\section*{Acknowledgments}
	
	This research was supported by the European Union's Horizon Europe Research and Innovation Programme under the Marie Sk\l{}odowska-Curie grant agreement No.~101149270--ProtoBH.
	
	\bibliographystyle{unsrtnat}
	\bibliography{2025_ICNFP_ROUPAS_arxiv}
	
\end{document}